\documentclass{desyprocA4}
\newcommand\pubdate{\today}

\def\to{\rightarrow}

\def\ph{\hat{P}}
\def\bi{\begin{itemize}}
 \def\ei{\end{itemize}}

\def\c1p{C1^\prime}
\def\msq3{\overline{m}_{\tilde{q}}(3)}
\def\ta{\tilde a}

\def\ta{\tilde a}

\def\tg{\tilde g}

\def\be{\begin{equation}}  
\def\ee{\end{equation}}  
\def\bea{\begin{eqnarray}}  
\def\eea{\end{eqnarray}}

\def\tz{\tilde Z}
\def\beq{\begin{equation}}
\def\eeq#1{\label{#1}\end{equation}}
\def\eeqn{\end{equation}}

\newenvironment{Eqnarray}%
   {\arraycolsep 0.14em\begin{eqnarray}}{\end{eqnarray}}
\def\beqa{\begin{Eqnarray}}
\def\eeqa#1{\label{#1}\end{Eqnarray}}
\def\eeqan{\end{Eqnarray}}

\begin{document}
%------------------------------------
\title{Implications of mixed axion/neutralino dark matter\\
for the Cosmic Frontier: a Snowmass whitepaper}

%for single authors the superscripts are optional
\author{{\slshape Kyu Jung Bae$^1$, Howard Baer$^1$ and Andre Lessa$^2$},\\
  $^1$Physics Dept., University of Oklahoma, Norman, OK 73019, USA\\ 
$^2$ Physics Dept. University of S\~ao Paulo, Brazil\\}
% please enter the contribution ID for the DOI
%\contribID{xy}

% TO THE CONFERENCE EDITORS: 
% please update the following information      
% before sending the template to the authors
%\confID{   }  % if the conference is on Indico uncomment this line
%\desyproc{ \quad }
%\acronym{PLHC2010} % if you want the Acronym in the page footer uncomment this line
%\doi  % if there is an online version we will register DOIs

\maketitle

\pubdate

\begin{abstract}
If one simultaneously invokes the SUSY solution to the gauge hierarchy problem and the
PQ solution to the strong $CP$ problem, then one might expect mixed axion/neutralino dark matter (DM),
{\it i.e.} two dark matter particles. In this case, dark matter production involves
oscillation production of axions, thermal and non-thermal neutralino production and 
thermal and non-thermal production of axinos and saxions. 
The latter particles may feed additional neutralinos
into the cosmic soup, or dilute all relics via additional entropy production, or feed dark radiation
into the cosmos. Depending on the SUSY and PQ model, the PQ scale can easily exist from
$10^9-10^{16}$ GeV. Axions may lie well beyond the range currently being probed by experiment.
In the case of relic WIMPs, theoretical predictions must be rescaled for the possibility that WIMPs
only comprise a fraction of the local DM abundance. 
\end{abstract}

% ****************************************************************************
%

\section{Introduction}

The Standard Model (SM) of particle physics is beset with two finetuning problems. The first-- the
well-known gauge hierarchy problem (GHP)-- 
pertains to quantum instability of scalar fields as manifested by quadratic divergences. 
The GHP can be solved by enlarging the spacetime symmetry group to include supersymmetry (SUSY). 
Realistic supersymmetric models invoke softly broken SUSY at the weak scale leading to the
prediction of a host of new states of matter: the so-called superpartners \cite{susyreview}.
The lightest SUSY particle (LSP) in $R$-parity conserving SUSY theories is absolutely stable 
and can be a dark matter (DM) candidate. In the case of a neutralino LSP, the DM particle would be a weakly interacting massive particle (WIMP)
with relic abundance from thermal freeze-out:
\begin{equation}
\Omega_{\chi}h^2\simeq\frac{3\times 10^{-27}\ {\rm cm^3/s}}{\langle\sigma_A v\rangle} .
\end{equation}
Relic WIMPs are being searched for in underground, space-based and ice-based detectors.

The second finetuning problem arises in quantum chromodynamics (QCD). The approximate global chiral symmetry of QCD 
$U(2)_L\times U(2)_R$ , arising from the presence of two light quark fields, 
provides an understanding of the origin of isospin symmetry and baryon number conservation. 
At the same time it naively predicts four instead of three light pions: the $U(1)_A$ problem. 
't Hooft's solution of the $U(1)_A$ problem via the QCD theta vacuum and instanton effects
requires an additional $CP$ violating term in the QCD Lagrangian:
\begin{equation}
{\cal L}_{QCD}\ni \theta \frac{g_s^2}{32\pi^2}G_{\mu\nu}^A\tilde{G}^{A\mu\nu} .
\end{equation}
Measurements of the neutron EDM imply that both this term along with another arising
from the quark mass matrix are very tiny, 
{\it i.e.} $\bar{\theta}\equiv \theta +Arg[det (M_q)]\lesssim 10^{-10}$ .
Peccei and Quinn (PQ) explained the absence of $\bar{\theta}$ by invoking an additional global
$U(1)_{PQ}$ symmetry along with an axion field $a$ \cite{pq}:
\begin{equation}
{\cal L}_{QCD}\ni \frac{g_s^2}{32\pi^2}\frac{a}{f_a}G_{\mu\nu}^A\tilde{G}^{A\mu\nu} .
\end{equation}
The potential energy develops a minimum at $a/f_a+\bar{\theta}=0$ and the offending term
dynamically vanishes. A consequence of this elegant mechanism is the existence of a
physical axion state with mass 
\begin{equation}
m_a\simeq 6.2\ {\rm \mu eV}\left(\frac{10^{12}\ {\rm GeV}}{f_a}\right)
\end{equation}
where the PQ scale $f_a\gtrsim 10^9$ GeV to avoid excessive cooling in supernovae.
While the axion can decay via $a\to\gamma\gamma$, its lifetime is longer than the age of the universe.
Thus, the axion is also a good CDM candidate.
A population of cold relic axions can be produced via coherent oscillations starting at
$T_{QCD}\sim 1$ GeV leading to a predicted abundance 
\begin{equation}
\Omega_ah^2\simeq 0.18 \theta^2\left(\frac{f_a}{10^{12}\ {\rm GeV}}\right)^{1.19}\left(\frac{\Lambda_{QCD}}{400\ {\rm MeV}}\right) .
\end{equation}
Relic QCD axions are being searched for in microwave cavity experiments such as ADMX \cite{admx}.

Usually the neutralino and axion are considered as dark matter candidates in separate scenarios.
However, in order to solve both finetuning problems in the SM, 
it is natural to invoke simultaneously the SUSY and PQ symmetries.
Thus, both neutralino and axion are co-existing dark matter candidates, 
and we can investigate the new possible scenarios of cosmology in the pre-Big Bang nucleosynthesis (pre-BBN) era and dark matter properties.

\section{Mixed axion/neutralino dark matter production in the early universe}

If one invokes the SUSY solution to the GHP and the PQ solution to the strong $CP$ problem, 
then the axion field must be elevated to a superfield. In four-component notation,
\begin{equation}
\hat{a}=(s+ia)/\sqrt{2}+i\sqrt{2}\bar{\theta}\tilde{a}+i\bar{\theta}\theta_L{\cal F}_a
\end{equation}
where $s$ is the $R$-even spin-0 {\it saxion}, $\ta$ is the $R$-odd spin-1/2 {\it axino}, ${\cal F}_a$
is an auxiliary field and in this case the $\theta$s are anticommuting superspace co-ordinates.
In gravity-mediated SUSY breaking models, both $s$ and $\ta$ are expected to receive soft masses $\sim m_{3/2}$.
Both $s$ and $\ta$ can be produced thermally in the early universe; 
since their interactions are suppressed by $1/f_a$, the $s$ and $\ta$ are quasi-long-lived, and in the case
where $\ta$ is the LSP, it is absolutely stable and can even be a component of CDM.
In the case where the neutralino is the LSP, one expects mixed axion/neutralino CDM, {\it i.e.} CDM composed of
two different particles. The interplay of $a$, $\ta$, $s$ and $\tz_1$ modifies the pre-BBN cosmological
history, and greatly impacts dark matter production rates. The early universe cosmology then depends on
exactly which SUSY axion model one invokes.
In the following, we will discuss two types of SUSY axion models and their physical characteristics.

\subsection{SUSY Kim-Shifman-Vainshtein-Zakharov (KSVZ) cosmology}

In the SUSY KSVZ approach\cite{ksvz}, one introduces vector-like heavy quarks with interaction
\begin{equation}
\int d^2\theta_L\lambda f_a e^{\hat{a}/f_a}\hat{Q}\hat{Q}^c
\end{equation}
leading to the axion supermultiplet coupling to MSSM particles via $QQ^c$ loops.
The derivative coupling of $s$ and $\ta$ to gluons leads to thermal production rates\cite{axprod} which depend on 
the re-heat temperature $T_R$:
\bea
\frac{\rho_s^{TP}}{s}& \simeq & 1.33\times 10^{-5} g_s^6\ln\frac{1.01}{g_s}\left(\frac{10^{12}\ {\rm GeV}}{f_a}\right)^2
\left(\frac{T_R}{10^8\ {\rm GeV}}\right)m_s ,\\
\frac{\rho_{\ta}^{TP}}{s}& \simeq & 0.9\times 10^{-5} g_s^6\ln\frac{3}{g_s}\left(\frac{10^{12}\ {\rm GeV}}{f_a}\right)^2
\left(\frac{T_R}{10^8\ {\rm GeV}}\right)m_{\ta} .
\eea
The saxion can also be produced via coherent oscillations (CO):
\be
\frac{\rho_s^{CO}}{s} \simeq  1.9\times 10^{-5} \left(\frac{min[T_R,T_s]}{10^8\ {\rm GeV}}\right)
\left(\frac{f_a}{10^{12}\ {\rm GeV}}\right)^2\left(\frac{s_0}{f_a}\right)^2 ,
\ee
which are expected to be dominant at large $f_a\gtrsim 10^{12}$ GeV.
Here $T_s$ is the temperature at which the saxion starts to oscillate.

Axinos may decay into modes such as $\tg g$ and $\tz_i\gamma$ thus augmenting the neutralino abundance if decays occur 
after neutralino freeze-out. For sufficient axino production rates, this can lead to neutralino re-annihilation at times 
later than freeze-out which also augments the neutralino abundance \cite{ckls,blrs}.
The axino decay products also inject entropy into the cosmic soup, thus diluting all relics present at the time of decay.
Saxions may decay to $gg$ and $\gamma\gamma$ thus creating entropy dilution. If they decay to SUSY particles
such as $\tg\tg$ or $\ta\ta$ then they will augment the neutralino abundance. They may also decay via $s\to aa$ thus
contributing to dark radiation. Dark radiation is constrained by the effective number of additional
neutrinos $\Delta N_{eff}$. Current data seem to prefer some slight excess, but conservatively\cite{planck}
\be
\Delta N_{eff}\lesssim 1.6 .
\ee
Saxions or axinos which decay at times later than BBN onset ($T\sim 1$ MeV) are likely to upset 
standard light element abundances as predicted by BBN; such scenarios are usually excluded.

\begin{figure}[htb]
  \begin{center}
\includegraphics[width=0.6\textwidth]{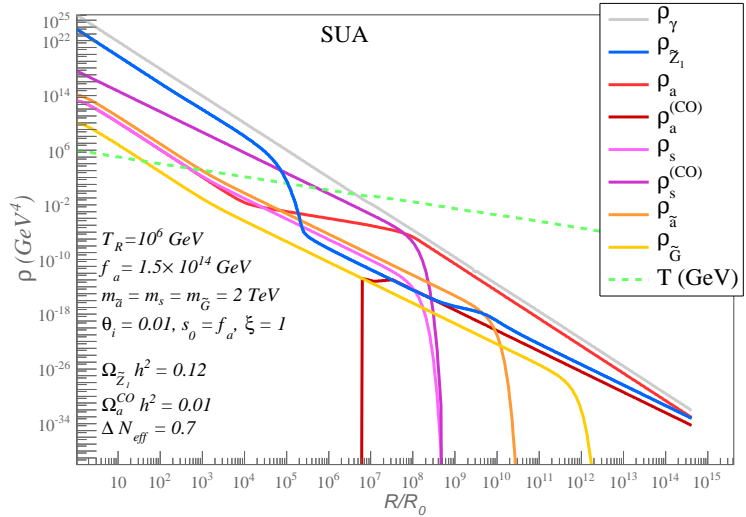}
  \end{center}
  \caption{Evolution of energy densities vs. scale factor for a SUSY model with
a standard  underabundance of neutralino dark matter (from Ref. \cite{bbl}).}
\label{fig:rho}
\end{figure}

The complex interplay of elements of the axion supermultiplet with neutralinos, gravitinos and radiation can be
computed using eight coupled Boltzmann equations. An illustration is provided in Fig. \ref{fig:rho}.
%
%%%%%%%%%%%%%%%%%%%%%%%%%%%%%%%%%%%%%%%%%%%%%%%%%%%%%%%%%%%%%%%%%%%%%%%%%%%%%%%%%%%%%%%
%%%%%%%%%%%%%%%%%%%%%%%%%%%%%%%%%%%%%%%%%%%%%%%%%%%%%%%%%%%%%%%%%%%%%%%%%%%%%%%%%%%%%%
%

In Ref. \cite{bbl}, it was found that SUSY models with a standard overabundance of neutralino dark matter
(as is typical for cases with a bino-like LSP) are typically still excluded under a PQMSSM cosmology with KSVZ axions.
At low $f_a\sim 10^9-10^{12}$ GeV, thermal production of axinos only augments the neutralino overabundance, while
at high $f_a\sim 10^{13}-10^{16}$ GeV, the neutralino and axion abundance can be entropy diluted but then one tends
to overproduce dark radiation from CO saxion production followed by $s\to aa$ decay. If $s\to aa$ decay is 
suppressed (as can occur in some models), then the large $f_a\sim 10^{13}-10^{16}$ GeV region can become allowed \cite{bl}
although one must be cautious to avoid BBN constraints. In this case, the DM abundance may be either
axion or neutralino dominated.

\begin{figure}[htb]
  \begin{center}
\includegraphics[width=0.7\textwidth]{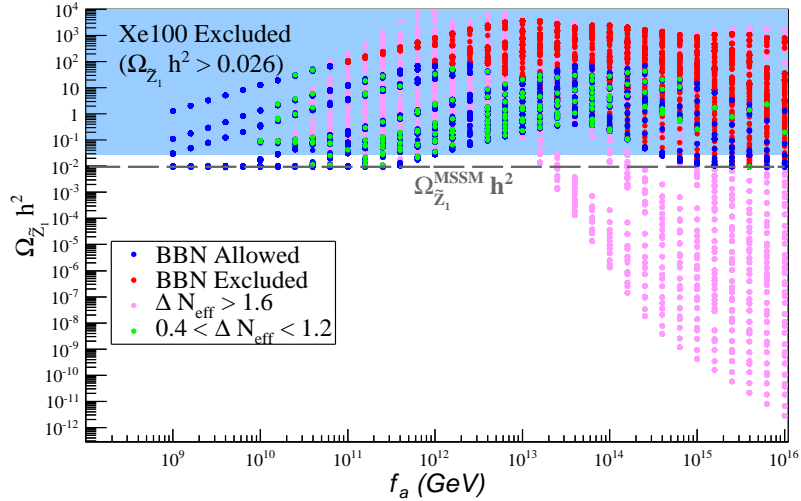}
  \end{center}
  \caption{Relic abundance of neutralinos in the PQMSSM with KSVZ axions for a SUSY model
with a standard  underabundance of higgsino-like neutralinos.
The green points accommodate a slight excess in dark radiation $\Delta N_{eff}$.}
\label{fig:SUA}
\end{figure}

For SUSY models with a standard underabundance of neutralinos, then a wide range of $f_a\sim 10^{9}-10^{13}$
can be allowed, and even values as high as $f_a\sim10^{15}-10^{16}$ are viable: see Fig. \ref{fig:SUA}.
%
%%%%%%%%%%%%%%%%%%%%%%%%%%%%%%%%%%%%%%%%%%%%%%%%%%%%%%%%%%%%%%%%%%%%%%%%%%%%%%%%%%%%%%%
%%%%%%%%%%%%%%%%%%%%%%%%%%%%%%%%%%%%%%%%%%%%%%%%%%%%%%%%%%%%%%%%%%%%%%%%%%%%%%%%%%%%%%
%

\subsection{SUSY Dine-Fischler-Srednicki- Zhitnitsky (DFSZ) cosmology}

The SUSY DFSZ model\cite{dfsz} is highly motivated in that it provides a solution to the supersymmetric $\mu$-problem \cite{knilles,chun}:
{\it i.e.} why is the superpotential higgsino mass $\mu\sim m_{weak}$ instead of $\sim m_{GUT}$?
One may invoke a superpotential of the form
\be
\hat{f}\ni \lambda_H\frac{\ph^2}{M_P}\hat{H}_u\hat{H}_d
\ee
where the singlet superfield $\ph$ carries PQ charge $+1$ and each Higgs superfield carries PQ charge $-1$.
After PQ symmetry breaking, then $\langle \hat{P}\rangle\simeq v_{PQ}=f_a/\sqrt{2}$ and the superpotential becomes
\be
\hat{f}\ni \mu\hat{H}_u\hat{H}_d
\ee
where now $\mu\simeq \lambda_H(v_{PQ}^2/M_P)$. With $v_{PQ}\sim 10^{10}$ GeV, then $\mu\sim 100$ GeV is 
generated, in accord with recent SUSY naturalness arguments \cite{rns}. As a byproduct, 
one also obtains a direct coupling of the axion superfield with the higgs multiplets:
\be
\hat{f}\ni 2\frac{\mu}{v_{PQ}}\hat{a}\hat{H}_u\hat{H}_d .
\ee

In the SUSY DFSZ model, the direct axion-higgs-higgs coupling leads to an axino
production rate roughly independent of $T_R$ \cite{bci,bci2}:
\be
Y_{\ta}^{TP}\simeq 10^{-5}\zeta\left(\frac{\mu}{{\rm TeV}}\right)^2\left(\frac{10^{11}\ {\rm GeV}}
{f_a}\right)^2 ,
\ee
where $\zeta \sim 1$ depends on the spectra.
Although a full calculation of thermal saxion production remains to be completed, one would expect
$Y_s^{TP}\sim Y_{\ta}^{TP}$. 

Also, since the axino and saxion directly couple to the Higgs multiplets, then the $\ta$ and $s$
decays tend to occur much sooner than expectations from the KSVZ case. Typically for the 
lower range of $f_a\sim 10^9-10^{12}$ GeV, one expects decay prior to neutralino freeze-out. 
In such cases, the standard thermal production rate for neutralino dark matter continues to hold, 
and cases of overproduction of neutralinos would be cosmologically excluded. Such a scenario
favors models with an standard underabundance of neutralinos, {\it i.e.} models with either
a wino-like or higgsino-like WIMP. Then the remaining abundance is filled by axions.
In these cases, one expects the CDM to be axion-dominated.

An exception to this situation occurs for large $f_a\sim 10^{13}-10^{16}$ GeV where thermal saxion
and axino production is suppressed, but where CO-production of saxions is enhanced.
In this case, depending on PQ parameters, saxion decay to SUSY particles will enhance the 
neutralino abundance, while decay to SM particles will dilute all relics and decay to axions will
add to and possibly exceed limits of dark radiation.

\section{Implications for dark matter detection}

In this Section we draw broad implications of the mixed axion/neutralino dark matter picture for
dark matter direct and indirect detection.

\subsection{Detection of axions}

In the case of axion detection, most of the current experimental effort has been focussed on searching for
relic axions in the few $\mu$eV range where the standard CO-production rate saturates the measured
abundance with misalignment angle $\theta_i\sim 1$. One implication of the SUSY scenario is that
{\it all of the measured dark matter abundance need not be axions}. Thus, the local abundance may only 
be a fraction of what is expected from an axion-only scenario. The second implication is that the
allowed range for $f_a$ can really range from very small values $f_a\sim 10^9$ GeV where the axion may
comprise only a fraction of the relic density all the way to $f_a\sim 10^{16}$ GeV where a 
combination of small misalignment angle plus entropy dilution still allows for axions to comprise the 
bulk of dark matter. No serious experiments have been proposed so far which can probe the
higher range of $f_a$ values. But even if axions are not found by ongoing searches, then the SUSY axion picture 
will not have suffered a significant blow to its allowed parameter space.

\subsection{Direct WIMP detection}

As with axions, the relic WIMP abundance will only be a fraction of its usually assumed value,
where one supposes $\rho_{DM}^{local}\sim 0.3\ {\rm GeV/cm^3}$. In the case of relic higgsino or wino-like
WIMPs, one must adjust theoretical predictions to (conservatively) account for the $\xi$ factor where
$\xi\equiv \Omega_{\chi}h^2/0.12$. For instance \cite{bbm}, the spin-independent direct detection rate
for higgsino-like WIMPs with $m_\chi \sim 100$ GeV is $\sigma^{SI}(\chi p)\sim 5\times 10^{-9}$ pb.
At first glance, these values seem excluded by recent Xe-100 searches \cite{Aprile:2012nq} which require
 $\sigma^{SI}(\chi p)\lesssim 3\times 10^{-9}$ pb. However, the local higgsino-like WIMP abundance
is typically a factor 10-15 lower than the measured value $\Omega_{CDM}h^2\simeq 0.12$, 
so the limits must be adjusted accordingly.

\subsection{Indirect WIMP detection}

Likewise, indirect WIMP detection limits must be adjusted for the possibility of WIMPs fulfilling 
only a portion of the local dark matter density. In the case of neutrino telescopes such as
IceCube, the neutrino flux from WIMP annihilation in the core of the sun depends more on the ability of 
the sun to sweep up WIMPs: if equilibration between capture and annihilation is attained, then 
no $\xi$ factor is needed. However, in the case of searches
for WIMP-WIMP annihilation in space into $\gamma$-rays or anti-matter, then theoretical projections must be multiplied by
a $\xi^2$ factor.

\section{Acknowledgments} This research was sponsored in part by grants from the US Department of Energy
and FAPESP.

\section{Bibliography}
% ****************************************************************************
% BIBLIOGRAPHY AREA
% ****************************************************************************

\begin{footnotesize}
% IF YOU DO NOT USE BIBTEX, USE THE FOLLOWING SAMPLE SCHEME FOR THE REFERENCES
% ----------------------------------------------------------------------------

\end{footnotesize}

% ****************************************************************************
% END OF BIBLIOGRAPHY AREA
% ****************************************************************************

\end{document}